\documentclass[a4paper,11pt]{article}
\usepackage{pos}
\usepackage{graphicx}
\usepackage{wrapfig,lipsum,booktabs}
\newcommand{\Tr}{{\rm Tr}}
\newcommand{\tr}{{\rm tr}}
\newcommand{\re}{{\rm Re}}

\newcommand{\Dodwf}{\mathcal{D}}

\title{Finite temperature QCD with physical $ (u/d, s, c)$ domain-wall quarks}
\ShortTitle{Finite temperature QCD with physical $ (u/d, s, c)$ domain-wall quarks}

\author[a,1]{Yu-Chih Chen}
\author*[a,b,c,1]{Ting-Wai Chiu}
\author[d,1]{Tung-Han Hsieh}

\affiliation[a]{
 Physics Department, National Taiwan University, Taipei, Taiwan 10617, Republic of China}

\affiliation[b]{
 Institute of Physics, Academia Sinica, Taipei, Taiwan 11529, Republic of China}

\affiliation[c]{
  Physics Department, National Taiwan Normal University, Taipei, Taiwan 11677, Republic of China}

\affiliation[d]{
Research Center for Applied Sciences, Academia Sinica, Taipei, Taiwan 11529, R.O.C.}

\emailAdd{twchiu@phys.ntu.edu.tw} 

\note{For the TWQCD Collaboration}

\abstract{
In order to understand the role of QCD in the early universe, 
we perform hybrid Monte-Carlo simulation of lattice QCD with 
$N_f=2+1+1$ optimal domain-wall quarks at the physical point, 
on the $64^3 \times (6,8,10,12,16,20,64)$ lattices,  
each with three lattice spacings.  
The lattice spacings and the bare quark masses are determined on the $64^4$ lattices. 
The resulting gauge ensembles provide a basis for studying finite temperature QCD 
with $N_f=2+1+1 $ domain-wall quarks at the physical point. In this Proceeding, 
we present our first result on the topological susceptibility of the QCD vacuum.
The topological charge of each gauge configuration 
is measured by the clover charge in the Wilson flow 
at the same flow time in physical units, 
and the topological susceptibility $ \chi_t(a,T) $ is determined for each ensemble 
with lattice spacing $a$ and temperature $T$.
Using the topological susceptibility $\chi_t(a,T) $ of 15 gauge ensembles   
with three lattice spacings and different temperatures in the range $T \sim 155-516 $~MeV,  
we extract the topological susceptibility $\chi_t(T)$ in the continuum limit.
}

\FullConference{%
 The 38th International Symposium on Lattice Field Theory, LATTICE2021
  26th-30th July, 2021
  Zoom/Gather@Massachusetts Institute of Technology
}

\begin{document}
\maketitle

\section{Introduction}

The topological susceptibility $ \chi_t $ is the most crucial quantity to measure
the quantum fluctuations of the QCD vaccum. 
Theoretically, the topological susceptibility is defined as 
\begin{equation}
\chi_t = \lim_{V \to \infty} \frac{\langle Q_t^2 \rangle}{V}, 
\end{equation}
where $ Q_t $ is the integer-valued topological charge of the gauge field 
in the 4-dimensional volume $V$, 
\begin{equation}
\label{eq:Qt}
Q_t=\frac{g^2 \epsilon_{\mu\nu\lambda\sigma}}{32 \pi^2} \int d^4 x \ \tr[ F_{\mu\nu}(x) F_{\lambda\sigma}(x)],
\end{equation}
and $ F_{\mu\nu} = T^a F_{\mu\nu}^a$ is the matrix-valued field tensor, with the normalization
$ \tr (T^a T^b) = \delta_{ab}/2 $.

At zero temperature,  
$ \chi_t $ is related to the chiral condensate $ \Sigma $,  
\begin{equation}
\Sigma = -\lim_{m_q \to 0} \lim_{V \to \infty} 
                           \frac{1}{V} \int d^4 x \ \langle \bar{q}(x) q(x) \rangle,  
\end{equation}
the order parameter of the spontaneously chiral symmetry breaking, and its nonzero value 
gives the majority of visible (non-dark) mass in the present universe.  

For QCD with $u$ and $d$ light quarks, the leading order chiral perturbation theory (ChPT) 
gives the relation \cite{Leutwyler:1992yt}
\begin{equation}
\label{eq:chit_sigma}
\chi_t = \Sigma \left( \frac{1}{m_u} + \frac{1}{m_d} \right)^{-1},  
\end{equation}  
which shows that $ \chi_t $ is proportional to $ \Sigma $.
This implies that the non-trivial topological quantum fluctuations in the QCD vacuum is 
the origin of the spontaneously chiral symmetry breaking. In other words,   
if $\chi_t $ is zero, then $ \Sigma $ is also zero and the chiral symmetry is unbroken, 
and the mass of the nucleon could be as light as $\sim 10 $~MeV rather than $\sim 940$~MeV.
Moreover, $\chi_t$ breaks the $ U_A(1) $ symmetry
and resolves the puzzle why the flavor-singlet $ \eta'$ is much
heavier than other non-singlet (approximate) Goldstone bosons 
\cite{tHooft:1976rip,Witten:1979vv,Veneziano:1979ec}.

At temperature $ T > T_c $, the chiral symmetry is restored and $ \Sigma = 0 $, 
thus the condition for deriving (\ref{eq:chit_sigma}) goes away, and the relation  
between $ \chi_t $ and $ \Sigma $ no longer holds. 
In other words, for $ T > T_c $, $ \chi_t $ and $ \Sigma $ are independent, 
thus the restoration of chiral symmetry does not necessarily implies the restoration 
of $U_A(1) $ symmetry. Interestingly, the non-trivial quantum fluctuations of the QCD vacuum 
at $ T > T_c $ only have the possibility to give a nonzero $ \chi_t $ but not the $ \Sigma $. 

For $T>T_c$, $ \chi_t(T)$ could play an important role 
in generating the majority of the mass in the universe,  
as a crucial input to the axion mass and energy density, a promising candidate 
for the dark matter in the universe. 
The axion 
\cite{Peccei:1977hh,Weinberg:1977ma,Wilczek:1977pj} 
is a pseudo Nambu-Goldstone boson arising from the breaking of a 
hypothetical global chiral U(1) extension of the Standard Model 
at an energy scale $f_A$ much higher than the electroweak scale, the Pecci-Quinn mechanism.  
This not only solves the strong CP problem, but also provides an explanation 
for the dark matter in the universe.  
The axion mass at temperature $T$ is proportional to $ \sqrt{\chi_t(T)} $,           
which is one of the key inputs to the equation of motion for the axion field evolving from 
the early universe to the present one, with solutions predicting the relic axion energy density,  
through the misalignment mechanism \cite{Dine:1981rt,Preskill:1982cy,Abbott:1982af}. 

For $ T < T_c $, the ChPT provides a prediction of $ \chi_t(T) $ 
with the input $ \chi_t(0) $ at the zero temperature 
\cite{Gasser:1986vb,Hansen:1990yg}.
However, for $ T > T_c $, the chiral symmetry is restored and the ChPT breaks down,  
thus the determination of $ \chi_t(T) $ requires a non-perturbative treatment 
from the first principles of QCD.  
To this end, lattice QCD provides a viable nonperturbative determination of $ \chi_t(T) $. 
Nevertheless, it becomes more and more challenging as the temperature gets higher and higher, 
since in principle the non-trivial configurations are more suppressed at higher temperatures, 
which in turn must require a much higher statatics in order to give a reliable determination. 
So far, direct simulations have only measured $ \chi_t(T) $ up to $ T \sim 550 $~MeV. 

Recent lattice studies of $\chi_t(T) $ aiming at the axion cosmology include 
various simulations with $N_f=0$, $2+1$, and $2+1+1$, 
where the lattice fermions in the unquenched simulations
include the staggered fermion, the Wilson fermion, and the twisted-mass Wilson fermion 
\cite{Berkowitz:2015aua, Kitano:2015fla, Borsanyi:2015cka, Bonati:2015vqz, 
      Petreczky:2016vrs, Borsanyi:2016ksw, Burger:2018fvb}.
For recent reviews, see, e.g., Refs. \cite{Moore:2017ond, Lombardo:2020bvn} 
and references therein.

In this study, we perform the HMC simulation of lattice QCD with $N_f=2+1+1$ 
optimal domain-wall quarks at the physical point, 
on the $64^3 \times (6, 8, 10, 12, 16, 20, 64)$ lattices, 
each with three lattice spacings $a \sim (0.064, 0.068, 0.075) $~fm. 
The bare quark masses and lattice spacings are determined on the $64^4$ lattices. 
The topological susceptibility of each gauge ensemble is measured by the Wilson flow 
at the flow time $ t = 0.8192~{\rm fm}^2 $, 
with the clover definition for the topological charge. 
Using the topological susceptibility $\chi_t(a,T)$ 
of 15 gauge ensembles with 3 different lattice spacings 
and different temperatures in the range $T \sim 155-516$~MeV, 
we extract the topological susceptibility $\chi_t(T)$ in the continuum limit.

\section{Gauge ensembles}

Our present simulations with physical $(u/d, s, c)$
on the $64^3 \times (6, 8, 10, 12, 16, 20, 64) \equiv N_x^3 \times N_t $ lattices 
are extensions of our previous ones \cite{Chen:2017kxr,Chiu:2018qcp,Chiu:2020ppa}, 
using the same actions and algorithms, and the same simulation code with tunings 
for the computational platform Nvidia DGX-V100.
Most of our production runs were performed on 10-20 units of Nvidia DGX-V100 at two institutions 
in Taiwan, namely, Academia Sinica Grid Computing (ASGC) and 
National Center for High Performance Computing (NCHC), from 2019 to 2021. 
Besides Nvidia DGX-V100, we also used other Nvidia GPU cards 
(e.g., GTX-2080Ti, GTX-1080Ti, GTX-TITAN-X, GTX-1080) for HMC simulations on 
the $64^3 \times (6,8,12) $ lattices, which only require 8-22 GB device memory. 
We outline our HMC simulations as follows.

For the gluon fields, we use the Wilson plaquette action 
\begin{equation*}
S_g(U) = \beta \sum_{\text{plaq.}} \left\{1-\frac{1}{3} \re \Tr (U_p) \right\},  
\end{equation*}
where $ \beta = 6/g_0^2 $. 
Then setting $\beta$ to three different values 
$\{ 6.15, 6.18, 6.20 \}$ gives three different lattice spacings.
%
For the quark fields, we use the optimal domain-wall fermion \cite{Chiu:2002ir}
and its extension with the $ R_5 $ symmetry \cite{Chiu:2015sea}. 
For domain-wall fermions, to simulate $ N_f = 2 +1 + 1 $
amounts to simulate $ N_f = 2 + 2 + 1 $ since 
\begin{eqnarray}
\label{eq:Nf2p2p1}
\left(\frac{\det \Dodwf(m_{u/d})}{\det \Dodwf(m_{PV})} \right)^2
\frac{\det \Dodwf(m_s)}{\det \Dodwf(m_{PV})}
\frac{\det \Dodwf(m_c)}{\det \Dodwf(m_{PV})}    
=
\left( \frac{\det \Dodwf(m_{u/d})}{\det \Dodwf(m_{PV})} \right)^2
\left( \frac{\det \Dodwf(m_c)}{\det \Dodwf(m_{PV})} \right)^2
\frac{\det \Dodwf(m_s)}{\det \Dodwf(m_{c})},
\end{eqnarray}
where $ \Dodwf(m_q) $ denotes the domain-wall fermion operator with bare quark mass $ m_q $, 
and $ m_{PV} $ is the Pauli-Villars mass. Since the simulation of 2-flavors
is more efficient than that of one-flavor, we use the RHS of (\ref{eq:Nf2p2p1}) 
for our HMC simulations. 
For the two-flavor factors, 
we use the $N_f=2$ pseudofermion actions \cite{Chiu:2011bm, Chen:2019wmn}.
For the one-flavor factor, 
we use the exact one-flavor pseudofermion action (EOFA) for DWF \cite{Chen:2014hyy}.
The parameters of the pseudofermion actions are fixed as follows.
For the $\Dodwf(m_q) $ defined in Eq. (2) of Ref. \cite{Chen:2014hyy},
we fix $ c = 1, d = 0 $, $ m_0 = 1.3 $, $ N_s = 16 $, 
$ \lambda_{max} = 6.20 $, and $\lambda_{min} = 0.05 $.
In the molecular dynamics, in order to enhance the efficiency, we use the Omelyan integrator, 
the Sexton-Weingarten multiple time scale method, and the mass preconditioning. 
The linear systems for computing the fermion forces and actions 
are solved by the conjugate gradient with mixed precision.  

\begin{wraptable}{r}{0.45 \textwidth}
\caption{The lattice parameters and statistics of the 15 gauge ensembles with $ T > T_c $.  
}
\begin{tabular}{cccccc}  
\toprule
    $\beta$
  & $a$[fm]
  & $ N_x $
  & $ N_t $
  & $T$[MeV]
  & $N_{\rm confs}$
\\                         \midrule 
6.20 & 0.0636 & 64 & 20 & 155 & 581   \\
6.18 & 0.0685 & 64 & 16 & 180 & 650   \\
6.20 & 0.0636 & 64 & 16 & 193 & 1577  \\
6.15 & 0.0748 & 64 & 12 & 219 & 566   \\
6.18 & 0.0685 & 64 & 12 & 240 & 500   \\
6.20 & 0.0636 & 64 & 12 & 258 & 1373  \\
6.15 & 0.0748 & 64 & 10 & 263 & 690   \\
6.18 & 0.0685 & 64 & 10 & 288 & 665   \\
6.20 & 0.0636 & 64 & 10 & 310 & 2547  \\
6.15 & 0.0748 & 64 & 8  & 329 & 1581  \\
6.18 & 0.0685 & 64 & 8  & 360 & 1822  \\
6.20 & 0.0636 & 64 & 8  & 387 & 2665  \\
6.15 & 0.0748 & 64 & 6  & 438 & 1714  \\
6.18 & 0.0685 & 64 & 6  & 479 & 1983  \\
6.20 & 0.0636 & 64 & 6  & 516 & 3038  \\
\bottomrule
\end{tabular}
\label{tab:15_ensembles}
\end{wraptable}

The initial thermalization of each ensemble is performed in one node with 1-8 GPUs interconnected 
by the NVLink. After thermalization, a set of gauge configurations are sampled and distributed 
to 8-16 simulation units, and each unit performs an independent stream of HMC simulation.  
Here one simulation unit consists of 1-8 GPUs in one node, 
depending on the size of the device memory and the computational efficiency.   
Then we sample one configuration every 5 trajectories in each stream, and obtain 
a total number of configurations for each ensemble. 
The statistics of the 15 gauge ensembles with $ T > T_c \sim 150 $~MeV are listed in
Table \ref{tab:15_ensembles}, where $T = 1/(N_t a) $.

The lattice spacings and bare quark masses are determined on the $64^4$ lattice.  
For the determination of the lattice spacing, 
we use the Wilson flow \cite{Narayanan:2006rf,Luscher:2010iy} with the condition 
\begin{equation*}
\label{eq:t0}
\left. \{ t^2 \langle E(t) \rangle \} \right|_{t=t_0} = 0.3, 
\end{equation*}
to obtain $\sqrt{t_0}/a $,  
then to use the input $ \sqrt{t_0} = 0.1416(8) $~fm \cite{Bazavov:2015yea} to 
obtain the lattice spacing $ a $. 
The lattice spacings for $\beta = \{6.15, 6.18, 6.20 \}$ are listed in Table \ref{tab:a_qmass}. 
In all cases, the spatial volume satisfies $ L^3 > (4~{\rm fm})^3 $ and $ M_\pi L \gtrsim 3 $.   

\begin{wraptable}{r}{0.45 \textwidth}
\caption{The lattice spacings and the bare quark masses of the gauge ensembles.}
\begin{tabular}{ccccc}
\toprule
    $\beta$ 
  & $a$[fm]
  & $ m_{u/d} a $
  & $ m_s a $
  & $ m_c a $ \\
\midrule
6.15 &  0.0748(1) &  0.00200 & 0.064 & 0.705 \\
6.18 &  0.0685(1) &  0.00180 & 0.058 & 0.626 \\
6.20 &  0.0636(1) &  0.00125 & 0.040 & 0.550 \\
\bottomrule
\end{tabular}
\label{tab:a_qmass}
\end{wraptable}

For each lattice spacing, the bare quark masses of $u/d$, $s$ and $c$ are tuned 
such that the lowest-lying masses of the meson operators 
$ \{ \bar{u} \gamma_5 d, \bar{s} \gamma_i s, \bar{c} \gamma_i c \} $ 
are in agreement with the physical masses of 
$\{ \pi^{\pm}(140), \phi(1020), J/\psi(3097) \} $ respectively. 
The bare quark masses of $u/d$, $s$, and $c$ of each lattice spacing 
are listed in Table \ref{tab:a_qmass}. 

To measure the chiral symmetry breaking due to finite $N_s=16$, we compute the residual mass
according to the formula derived in Ref. \cite{Chen:2012jya}. 
The residual masses of $u/d$, $s$, and $c$ quarks are computed  
for each of the 15 ensembles in this study, and they are   
less than $1.86\%$, $0.05\%$ and $0.002\%$ of their bare masses respectively. 
In the units of MeV/$c^2$, the residual masses 
of $u/d$, $s$ and $c$ quarks are less than 0.09, 0.08, and 0.04 respectively. 
This asserts that the chiral symmetry is well preserved such that the deviation 
of the bare quark mass $m_q$ is sufficiently small
in the effective 4D Dirac operator $ (D_c + m_q )/(1+rD_c) $ 
of the optimal domain-wall fermion, for both light and heavy quarks. 
In other words, the chiral symmetry in our simulations should be sufficiently precise 
to guarantee that the hadronic observables 
can be determined with a good precision, with the associated uncertainty 
much less than those due to statistics and other systematic ones.

\section{Topological charge and topological susceptibility}

\begin{wrapfigure}{r}{0.55 \textwidth}
\centering
\includegraphics[width=8cm,clip=true]{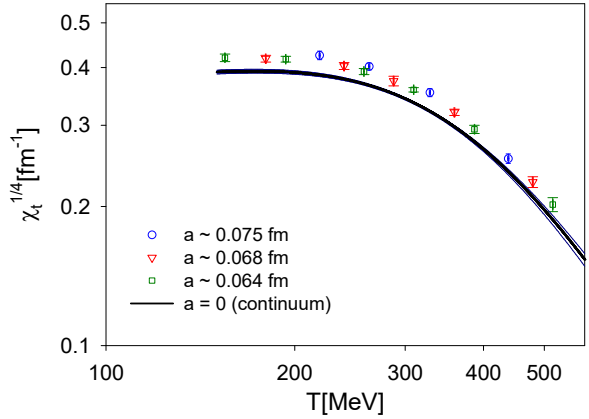}
\caption{
The fourth root of topological susceptibility $\chi_t^{1/4}$ versus the temperature $T$.
}
\label{fig:chit14_15pts_a0}
\end{wrapfigure} 
%
The topological charge $Q_t$ of each configuration is 
measured by the Wilson flow, using the clover definition.
The Wilson flow equation is integrated from the flow time $t/a^2=0$ to 256
with the step size $0.01$. 
In order to extrapolate the topological susceptibility $ \chi_t = \langle Q_t^2 \rangle/V $ 
to the continuum limit, $Q_t$ is required to be measured at the same physical flow time 
for each configuration, which is chosen to be $0.8192~{\rm fm}^2 $ such that $ \chi_t $ 
attains a plateau for each ensemble in this study. 

The results of $\chi_t^{1/4}(a,T)$ 
of 15 gauge ensembles are plotted in Fig. \ref{fig:chit14_15pts_a0}, 
which are denoted by blue circles ($a \sim 0.075$~fm), 
red inverted triangles ($a \sim 0.068$~fm), and green squares ($ a \sim 0.064 $~fm). 
%
First, we observe that the 5 data points of $ \chi_t^{1/4} $ at high temperature
$ T > 350 $~MeV can be fitted by the power law 
$ \chi_t^{1/4}(T) \sim T^{-p} $, independent of the lattice spacing $a$.
However, the power law cannot fit all 15 data points. 
In order to construct an analytic formula which can fit all data points of $ \chi_t(T) $ 
for all temperatures, one considers a function which behaves like 
the power law $ \sim (T_c/T)^p $ for $T \gg T_c $, but in general it  
incorporates all higher order corrections, i.e.,  
\begin{equation}
\label{eq:basic_idea}
\chi_t^{1/4}(T) = c_0 (T_c/T)^{p} 
\left[ 1 + b_1 (T_c/T) + b_2 (T_c/T)^2 + \cdot + b_n (T_c/T)^n + \cdots \right]. 
\end{equation}
In practice, it is vital to recast (\ref{eq:basic_idea})  
into a formula with fewer parameters, e.g.,  
\begin{equation}
\label{eq:a=0}
\chi_t^{1/4}(T) = c_0 \frac{(T_c/T)^{p}}{1+b_1(T_c/T) + b_2 (T_c/T)^2}.  
\end{equation}
It turns out that the 6 data points of $\chi_t^{1/4} $ at $ a \sim 0.064$~fm ($\beta = 6.20 $) 
are well fitted by (\ref{eq:a=0}).   
Thus, for the global fitting of all $ \chi_t^{1/4}(a,T) $ with different $a$ and $T$,
the simplest extension of (\ref{eq:a=0}) is to replace $ c_0 $ with $ (c_0 + c_1 a^2) $. 
This leads to the ansatz
\begin{equation}
\label{eq:ansatz}
\chi_t^{1/4}(a,T)= 
(c_0 + c_1 a^2) \frac{(T_c/T)^p}{1 + b_1 (T_c/T) + b_2 (T_c/T)^2}, \hspace{4mm} T_c = 150~{\rm MeV}. 
\end{equation}
Fitting the 15 data points of $ \chi_t^{1/4} $ in Fig. \ref{fig:chit14_15pts_a0} 
to (\ref{eq:ansatz}), it gives 
$c_0 = 1.89(3)$,  
$c_1 = 32.2(6.8)$, 
$p = 2.03(5)$, 
$b_1 = -2.42(19)$, 
$b_2 = 6.25(14)$ 
%
%
with $\chi^2$/d.o.f. = 0.21.
Note that the fitted value of the exponent $p$ is rather insensitive 
to the choice of $ T_c = 150~{\rm MeV} $, 
i.e., any value of $ T_c $ in the range of 145-155 MeV gives almost the same value of $p$.
Then $\chi_t^{1/4}(T)$ in the continuum limit can be obtained by setting $a^2 =0 $ 
in (\ref{eq:ansatz}), which is plotted as the solid black line in Fig. \ref{fig:chit14_15pts_a0}, 
with the error bars as the enveloping blue solid lines.
In the limit $ T \gg T_c $, it becomes 
$\chi_t^{1/4}(T) = c_0 (T_c/T)^{2.03(5)} $, i.e., $\chi_t(T) = c_0^4 (T_c/T)^{8.1(2)} $,     
which agrees with the temperature dependence of $\chi_t(T) $ in  
the dilute instanton gas approximation (DIGA) \cite{Gross:1980br}, i.e.,  
$ \chi_t(T) \sim T^{-8.3} $ for $N_f=4$. 
This also implies that our data points of $ \chi_t(a, T) $ (for $ T > 350 $~MeV) are valid, 
up to an overall constant factor. 

It is interesting to note that our 15 data points of $\chi_t(a, T) $ 
are only up to the temperature $ T \sim 515 $~MeV.   
Nevertheless, they are sufficient to fix the coefficents of (\ref{eq:ansatz}), 
which in turn can give $ \chi_t(T) $ for any $ T > T_c $. 
This is the major advantage of having an analytic formula like (\ref{eq:ansatz}). 
There are many possible variations of (\ref{eq:ansatz}), e.g., replacing $(c_0 + c_1 a^2) $ 
with $ (c_0 + c_1 a^2 + c_2 a^4) $, adding the $a^2$ term to the exponent $ p $ 
and/or the coefficients $ b_1 $ and $ b_2 $, etc.    
For our 15 data points, all variations give consistent results of $ \chi_t(T) $ 
in the continuum limit.

\section{Discussions}

To summarize, this is the first determination of $ \chi_t(T) $ in lattice QCD
with $N_f = 2+1+1 $ optimal domain-wall quarks at the physical point, by direct simulations.
Here the chiral symmetry is preserved with $N_s=16$ in the fifth dimension, 
and the optimal weights $\{\omega_s, s=1,\cdots, 16\}$ are computed with 
$\lambda_{min} = 0.05 $ and $ \lambda_{max} = 6.2 $, and the error 
of the sign function of $H_w$ is less than $1.2 \times 10^{-5} $, for eigenvalues of $ H_w $ 
satisfying $ \lambda_{min} \le | \lambda(H_w) | \le \lambda_{max} $. 
However, it is not in the exact chiral symmetry limit,  
the smallest eigenvalue of 
the effective 4D Dirac operator $ D(m_q)=(D_c + m_q )/(1+rD_c) $ is larger than $ m_q $. 
Thus the fermion determinant is larger than its value in the exact chiral symmetry limit. 
Now the question is how $ \chi_t(T) $ depends on the chiral symmetry in this study. 
For optimal domain-wall fermion, the exact chiral symmetry is in the limit  
$ N_s \to \infty $ and $ \lambda_{min} \to 0 $.  
In practice, this can be attained by increasing $N_s$ and 
decreasing $ \lambda_{min} $ such that the error due to the chiral symmetry breaking becomes
negligible in any physical observables. For example, if one takes 
$ N_s = 32 $, $\lambda_{min} = 10^{-4} $ and $ \lambda_{max} = 6.2 $, then the error  
of the sign function of $H_w$ is less than $1.2 \times 10^{-5} $ for eigenvalues of $ H_w $ 
satisfying $ 10^{-4} \le | \lambda(H_w) | \le 6.2 $. 
Nevertheless, this set of simulations is estimated to be $ \sim 100$ times
more expensive than the present one, beyond the limit of our present resources.
At this point, one may wonder whether it is possible to use 
the reweighting method to obtain $ \chi_t(a,T) $ in the exact chiral symmetry limit, 
without performing new simulations at all. 
However, according to our discussion of the reweighting method for DWF \cite{Chen:2022fid}, 
it is infeasible to apply the reweighting method to the $ \chi_t(a,T) $ results in 
the present study, thus new simulations with smaller $ \lambda_{min} $ and larger $ N_s $ 
are required.

\section*{Acknowledgement}

We are grateful to Academia Sinica Grid Computing Center (ASGC)
and National Center for High Performance Computing (NCHC) for the computer time and facilities.
This work is supported by the Ministry of Science and Technology
(Grant Nos.~108-2112-M-003-005, 109-2112-M-003-006, 110-2112-M-003-009).


\begin{thebibliography}{99}

\bibitem{Leutwyler:1992yt}
H.~Leutwyler and A.~V.~Smilga,
Phys. Rev. D \textbf{46}, 5607-5632 (1992)

\bibitem{tHooft:1976rip}
G.~'t Hooft,
Phys. Rev. Lett. \textbf{37}, 8-11 (1976); 
%
Phys. Rev. D \textbf{14}, 3432-3450 (1976)
[erratum: Phys. Rev. D \textbf{18}, 2199 (1978)]

\bibitem{Witten:1979vv}
E.~Witten,
Nucl. Phys. B \textbf{156}, 269-283 (1979)

\bibitem{Veneziano:1979ec}
G.~Veneziano,
Nucl. Phys. B \textbf{159}, 213-224 (1979)

\bibitem{Peccei:1977hh}
R.~D.~Peccei and H.~R.~Quinn,
Phys. Rev. Lett. \textbf{38}, 1440-1443 (1977); 
%
Phys. Rev. D \textbf{16}, 1791-1797 (1977)

\bibitem{Weinberg:1977ma}
S.~Weinberg,
Phys. Rev. Lett. \textbf{40}, 223-226 (1978)

\bibitem{Wilczek:1977pj}
F.~Wilczek,
Phys. Rev. Lett. \textbf{40}, 279-282 (1978)

\bibitem{Dine:1981rt}
M.~Dine, W.~Fischler and M.~Srednicki,
Phys. Lett. B \textbf{104}, 199-202 (1981)

\bibitem{Preskill:1982cy}
J.~Preskill, M.~B.~Wise and F.~Wilczek,
Phys. Lett. B \textbf{120}, 127-132 (1983)

\bibitem{Abbott:1982af}
L.~F.~Abbott and P.~Sikivie,
Phys. Lett. B \textbf{120}, 133-136 (1983)

\bibitem{Gasser:1986vb}
J.~Gasser and H.~Leutwyler,
Phys. Lett. B \textbf{184}, 83-88 (1987)





\bibitem{Hansen:1990yg}
F.~C.~Hansen and H.~Leutwyler,
Nucl. Phys. B \textbf{350}, 201-227 (1991)



\bibitem{Berkowitz:2015aua}
E.~Berkowitz, M.~I.~Buchoff and E.~Rinaldi,
Phys. Rev. D \textbf{92}, no.3, 034507 (2015)
[arXiv:1505.07455 [hep-ph]].

\bibitem{Kitano:2015fla}
R.~Kitano and N.~Yamada,
JHEP \textbf{10}, 136 (2015)
[arXiv:1506.00370 [hep-ph]].

\bibitem{Borsanyi:2015cka}
S.~Borsanyi, M.~Dierigl, Z.~Fodor, S.~D.~Katz, S.~W.~Mages, D.~Nogradi, J.~Redondo, A.~Ringwald and K.~K.~Szabo,
Phys. Lett. B \textbf{752}, 175-181 (2016)
[arXiv:1508.06917 [hep-lat]].

\bibitem{Bonati:2015vqz}
C.~Bonati, M.~D'Elia, M.~Mariti, G.~Martinelli, M.~Mesiti, F.~Negro, F.~Sanfilippo and G.~Villadoro,
JHEP \textbf{03}, 155 (2016)
[arXiv:1512.06746 [hep-lat]].

\bibitem{Petreczky:2016vrs}
P.~Petreczky, H.~P.~Schadler and S.~Sharma,
Phys. Lett. B \textbf{762}, 498-505 (2016)
[arXiv:1606.03145 [hep-lat]].

\bibitem{Borsanyi:2016ksw}
S.~Borsanyi, Z.~Fodor, J.~Guenther, K.~H.~Kampert, S.~D.~Katz, T.~Kawanai, T.~G.~Kovacs, S.~W.~Mages, A.~Pasztor and F.~Pittler, \textit{et al.}
Nature \textbf{539}, no.7627, 69-71 (2016)
[arXiv:1606.07494 [hep-lat]].

\bibitem{Burger:2018fvb}
F.~Burger, E.~M.~Ilgenfritz, M.~P.~Lombardo and A.~Trunin,
Phys. Rev. D \textbf{98}, no.9, 094501 (2018)
[arXiv:1805.06001 [hep-lat]].

\bibitem{Moore:2017ond}
G.~D.~Moore,
EPJ Web Conf. \textbf{175}, 01009 (2018)
[arXiv:1709.09466 [hep-ph]].

\bibitem{Lombardo:2020bvn}
M.~P.~Lombardo and A.~Trunin,
Int. J. Mod. Phys. A \textbf{35}, no.20, 2030010 (2020)
[arXiv:2005.06547 [hep-lat]].


%

\bibitem{Chen:2017kxr}
  Y.~C.~Chen and T.~W.~Chiu [TWQCD Collaboration],
  Phys.\ Lett.\ B {\bf 767}, 193 (2017)
  [arXiv:1701.02581 [hep-lat]].


\bibitem{Chiu:2018qcp}
T.~W.~Chiu [TWQCD Collaboration],
PoS \textbf{LATTICE2018}, 040 (2018)
[arXiv:1811.08095 [hep-lat]].


\bibitem{Chiu:2020ppa}
T.~W.~Chiu [TWQCD Collaboration],
PoS \textbf{LATTICE2019}, 133 (2020)
[arXiv:2002.06126 [hep-lat]].



\bibitem{Chiu:2002ir}
  T.~W.~Chiu,
  Phys.\ Rev.\ Lett.\  {\bf 90}, 071601 (2003)
  [hep-lat/0209153] 

\bibitem{Chiu:2015sea}
  T.~W.~Chiu,
  Phys.\ Lett.\ B {\bf 744}, 95 (2015)
  [arXiv:1503.01750 [hep-lat]].


\bibitem{Chiu:2011bm}
  T.~W.~Chiu, T.~H.~Hsieh, Y.~Y.~Mao [TWQCD Collaboration],
  Phys.\ Lett.\ B {\bf 717}, 420 (2012)
  [arXiv:1109.3675 [hep-lat]].

\bibitem{Chen:2019wmn}
Y.~C.~Chen and T.~W.~Chiu,
Phys. Rev. D \textbf{100}, no.5, 054513 (2019)
[arXiv:1907.03212 [hep-lat]].

\bibitem{Chen:2014hyy}
  Y.~C.~Chen and T.~W.~Chiu [TWQCD Collaboration],
  Phys.\ Lett.\ B {\bf 738}, 55 (2014)
  [arXiv:1403.1683 [hep-lat]].



\bibitem{Narayanan:2006rf}
  R.~Narayanan and H.~Neuberger,
  JHEP {\bf 0603}, 064 (2006)
  [hep-th/0601210].

\bibitem{Luscher:2010iy}
  M.~Luscher,
  JHEP {\bf 1008}, 071 (2010)
  Erratum: [JHEP {\bf 1403}, 092 (2014)]
  [arXiv:1006.4518 [hep-lat]].

\bibitem{Bazavov:2015yea}
  A.~Bazavov {\it et al.} [MILC Collaboration],
  Phys.\ Rev.\ D {\bf 93}, no. 9, 094510 (2016)
  [arXiv:1503.02769 [hep-lat]].

\bibitem{Chen:2012jya}
  Y.~C.~Chen, T.~W.~Chiu [TWQCD Collaboration],
  Phys.\ Rev.\ D {\bf 86}, 094508 (2012)
  [arXiv:1205.6151 [hep-lat]].



\bibitem{Gross:1980br}
D.~J.~Gross, R.~D.~Pisarski and L.~G.~Yaffe,
Rev. Mod. Phys. \textbf{53}, 43 (1981)



\bibitem{Chen:2022fid}
Y.~C.~Chen, T.~W.~Chiu and T.~H.~Hsieh,
[arXiv:2204.01556 [hep-lat]].


\end{thebibliography}
\end{document}